# DEVELOPING A GENERAL ALGORITHM FOR BALL CURVE WITH $GC^2$

**[§]Gobithasan, R., [†]Norziah, O., [◊]Jamaludin M.A**
*[§] Dept. of Mathematics, FST, KUSTEM, Mengabang Telipot,*
*21030, Kuala Terengganu, M'sia.*
*Tel: +609-668 3132. Fax: +609-669 4660,*
*email: gobithasan@kustem.edu.my*
*[§,◊,†] School of Mathematical Sciences, Universiti Sains Malaysia,*
*11800 Minden, Penang, M'sia.*
*Tel: +604-653 3656. Fax: +604-657 0910,*
*email: nziah@yahoo.com, jamaluma@cs.usm.my*

**Abstract:** This paper dwells in developing a general algorithm for constructing a piecewise Ball Curve with curvature continuity ($GC^2$). The proposed algorithm requires $GC^2$ data in which the designer must define unit tangent vectors and signed curvatures at each interpolating points. As a numerical example, a vase is constructed using $GC^2$ piecewise Ball Curve.

## 1. Introduction

There are many polynomial bases used to define curves namely Bezier, Overhauser, Timmer and Ball. Ball curve was first introduced by Alan Ball [1,2,3] who was working for British Aircraft Corporation, BAC, and it was used for BAC's in-house CAD system.

Geometric continuity, GC, is one of the fields in computer aided geometric design (CAGD) which is constantly researched. Designing $GC^2$ curves and surfaces for computer aided design and manufacturing CAD/CAM systems has been a tough problem to deal with for a long time. Early efforts can be seen in [4,5,6]. There are also works has been done for developing composite $GC^2$ Timmer curve [7,8,9]. In addition to prior efforts aforesaid, this paper proposes a customized algorithm to construct a $GC^2$ continuous Ball curve.

Ball bases indicate two advantages when compared to Bezier curves. Firstly, there exists a robust time effective algorithm to evaluate the Ball Curve as compared to the evaluating Bezier Curve using the Casteljau algorithm [10]. Secondly, generalized Ball bases are said to suit much better in degree elevation and reduction [11]. This point is imperative when it comes to data transfer among Computer Aided Design (CAD) systems. To add, it also can handle conic sections in the form of rational cubic. Thus, the developed algorithm can be utilized for product design with numerical controls, NCs, whereby the curvature value at the interpolating points are controllable.

The proposed scheme is an interactive algorithm in which the designers need to define $G^2$ data for respective interpolating points (data points). In the last section, the advantages of the proposed algorithm are elaborated with regards of a vase as the numerical example.

## 2. Ball Cubic

Let $P_i(x_i, y_i)$, $P_{i,1}(x_{i,1}, y_{i,1})$, $P_{i,2}(x_{i,2}, y_{i,2})$ and $P_{i+1}(x_{i+1}, y_{i+1})$ be represented as the control points of a Ball curve, $B_i(t)$. Thus, Ball parametric cubic can be written as [1,2,3]:

$$B_i(t) = (x(t), y(t)) = (1-t)^2 P_i + 2t(1-t)^2 P_{i,1} + 2t^2(1-t)P_{i,2} + t^2 P_{i+1} \quad (1)$$

where $0 \le t \le 1$. The alternative way of rendering a Ball curve which is undertaken in this research is by using the interpolating points, $P_i$ and $P_{i+1}$, and its respective unit tangent vectors, $T_i(M_i, N_i)$ and $T_{i+1}(M_{i+1}, N_{i+1})$. By utilizing Eq. (2) and Eq. (3), the intermediate points $P_{i,1}$ and $P_{i,2}$ are calculated:

$$\begin{aligned} P_{i,1}(x_{i,1}, y_{i,1}) &= P_i + \frac{1}{\alpha_i} T_i \\ &= (x_i + \frac{1}{\alpha_i} M_i, y_i + \frac{1}{\alpha_i} N_i) \end{aligned} \quad (2)$$

$$\begin{aligned} P_{i,2}(x_{i,2}, y_{i,2}) &= P_{i+1} - \frac{1}{\beta_i} T_{i+1} \\ &= (x_{i+1} - \frac{1}{\beta_i} M_{i+1}, y_{i+1} - \frac{1}{\beta_i} N_{i+1}) \end{aligned} \quad (3)$$

where $\alpha_i$ and $\beta_i$ are real positive numbers. The advantage of this approach is that one may control the direction of the Ball curve by only controlling the given value of unit tangent vector.

The properties of Ball curve are:

- coordinate system independent,
- obeys convex hull property, CHP
- obeys variation diminishing property, VHP
- symmetry,
- it takes invariant form under affine transformation, and
- it interpolates the end points.

## 3. An Interactive Algorithm to Generate A $G^2$ Ball Curve

$G^2$ data is a set of points, unit tangent vectors and signed curvatures at the respective points. In $G^2$ interpolation, the curve passes through the given points, matches given unit tangent vectors and signed curvatures at the respective points [4].

Let a Ball curve $B_i(t)$ with its control points and unit tangent vectors be denoted as in the first section. One needs to find a suitable pair of $\alpha_i$ and $\beta_i$ before calculating $P_{i,1}$ and $P_{i,2}$. By substituting Eq. (2) and Eq. (3) into Eq. (1), the Ball curve is now represented as:

$$B_i(t) = \frac{2(-1+t)(T_i(-1+3t) + 3t(P_i - P_{i+1})\alpha}{\alpha} + \frac{2T_{i+1}t(-2+3t)}{\beta} \quad (4)$$

The formula of signed curvature is defined in plane as [12]:

$$\begin{aligned} \kappa_i(t) &= \frac{B_i'(t) \times B_i''(t)}{\| B_i'(t) \|^3} \\ &= \frac{(x'(t), y'(t)) \times (x''(t), y''(t))}{(\sqrt{(x'(t))^2 + (y'(t))^2})^3} \\ &= \frac{(x'(t)y''(t)) - (x''(t)y'(t))}{((x'(t))^2 + (y'(t))^2)^{3/2}} \end{aligned} \quad (5)$$

The radius of the signed curvature, $R_i$, is the absolute value of the reciprocal of Eq. (5). Consequently, by substituting the stated $G^2$ data for $H_i(t)$ into Eq. (5), $\kappa_i(t)$ becomes:

$$\kappa_i(t) = \frac{\begin{array}{l} -4\{2M_iN_{i+1} - 2M_{i+1}N_i[1+3t(t-1)] - 3M_{i+1}t^2(y_i - y_{i+1})\alpha_i \\ +3N_{i+1}t[2M_i(t-1) + t(x_i - x_{i+1})\alpha_i] - 3(t-1)^2[N_i(x_i - x_{i+1}) + M_i(y_{i+1} - y_i)]\beta_i\} \end{array}}{\begin{array}{l} \alpha_i\{\{\frac{2(t-1)[M_i(3t-1)+3t(x_i - x_{i+1})\alpha_i]}{\alpha_i} + \frac{2M_it(3t-2)}{\beta_i}\}^2 \\ +\{\frac{2(t-1)[N_i(3t-1)+3t(y_i - y_{i+1})\alpha_i]}{\alpha_i} + \frac{4N_{i+1}t(3t-2)}{\beta_i}\}^2\}^{3/2}\beta_i \end{array}} \quad (6)$$

At t=0, $\kappa_i(t)$ becomes:

$$\kappa_i(0) = \frac{-4(-2M_{i+1}N_i + 2M_iN_{i+1} - 3[N_i(x_i - x_{i+1}) + M_i(y_{i+1} - y_i)\beta_i]\}}{(\frac{4M_i^2 + 4N_i^2}{\alpha_i^2})^{3/2}\alpha_i\beta_i} \quad (7)$$

Since $(M_i^2 + N_i^2) = 1$, thus one may simplify the denominator of Eq. (7) as stated below:

$$(\frac{4M_i^2 + 4N_i^2}{\alpha_i^2})^{3/2}\alpha_i\beta_i = \frac{8\beta_i}{\alpha_i^2} \quad (8)$$

Therefore Eq. (7) can be represented as Eq. (9) upon algebraic simplification:

$$\kappa_i(0) = \frac{\alpha^2\{2M_{i+1}N_i - 2M_iN_{i+1} + 3[N_i(x_i - x_{i+1}) + M_i(y_{i+1} - y_i)\beta_i]\}}{2\beta_i} \quad (9)$$

The signed curvature in term of radius at t=0, $R_i^0$ is:

$$\kappa_i(0) = S_i^0\frac{1}{R_i^0} \quad (10)$$

where $S_i^0$ indicates the sign of the curvature at $H_i(0)$. The curvature is given a positive sign if the circle of the curvature is on the left of the curve and a negative sign if the circle of the curvature is on the right of the curve [4]. Hence, Eq. (11) is obtained from Eq. (9) and (10).

$$S_i^0\frac{1}{R_i^0} = \frac{\alpha^2\{2M_{i+1}N_i - 2M_iN_{i+1} + 3[N_i(x_i - x_{i+1}) + M_i(y_{i+1} - y_i)\beta_i]\}}{2\beta_i} \quad (11)$$

By implementing the same method as described before, the signed curvature at t=1 is obtained:

$$\kappa_i(1) = \frac{\beta_i^2\{N_{i+1}[3(x_{i+1} - x_i)\alpha_i - 2M_i] + M_{i+1}[3(y_i - y_{i+1})\alpha + 2N_{ii}]\}}{2\alpha_i} \quad (12)$$

and

$$\kappa_i(1) = S_i^1\frac{1}{R_i^1} \quad (13)$$

By substituting Eq. (12) into Eq. (13), Eq. (14) is obtained:

$$S_i^1\frac{1}{R_i^1} = \frac{\beta^2\{N_{i+1}[3(x_{i+1} - x_i)\alpha_i - 2M_i] + M_{i+1}[3(y_i - y_{i+1})\alpha + 2N_i]\}}{2\alpha} \quad (14)$$

As a result, one may now solve the simultaneous equations (Eq. (11) and (14)) which are in the form of fourth degree equation in order to obtain the suitable pair of $\alpha_i$ and $\beta_i$.

Subsequently, one may obtain a suitable pair of $\alpha_i$ and $\beta_i$ by interactively setting the appropriate signs for curvatures at the interpolating points along with its suitable unit tangent vectors. The general algorithm is shown below:

(1). define $P_0$ and $P_1$ ;

define $T_0$ and $T_1$ ;

define the radius, $R_0^0$ and the sign, $S_0^0$ (at t=0 );

define the radius, $R_0^1$ and the sign, $S_0^1$ (at t=1 );

solve the simultaneous equations ( Eq. (11) and Eq. (14) ) to obtain $\alpha_0$ and $\beta_0$ ;

select a pair of $\alpha_0$ and $\beta_0$ with real positive numbers ;

render the curve, $H_0(t)$, with $0\leq t \leq 1$ where $H_0(t)$ is given by Eq. (4) ;

(2). **For** i=1 **to** n-1 **do**

define $P_{i+1}$ ;

define $T_{i+1}$ ;

$R_i^0 = R_{i-1}^1$ and $S_i^0 = S_{i-1}^0$ (at t=0 );

define the radius, $R_i^1$ and the sign, $S_i^1$ (at t=1 );

solve the simultaneous equation ( Eq. (11) and Eq. (14) ) to obtain $\alpha_i$ and $\beta_i$ ;

select a pair of $\alpha_i$ and $\beta_i$ with real positive number ;

render the curve, $H_i(t)$, with $0\leq t \leq 1$ where $H_i(t)$ is given by Eq. (4) ;

## 4. Numerical Example: A Profile of A Vase

In this section, four interpolating points are joined with $G^2$ continuity in order to obtain a profile of a vase. The points are: $P_0 = (1,0), P_1 = (3.5,5), P_2 = (0.5,9)$ and $P_3 = (2,12)$. The proposed algorithm stated in this paper is implemented in order to identify its strength. The profile of the vase is then revolved around the y- axis in order to obtain the surface of the vase.

By implementing the interactive $G^2$ algorithm, a different shape of $G^2$ piecewise Ball curve can be obtained. Since the designer is free to define the unit tangent vectors and signed curvatures are the interpolating points, thus the curve can be customized as desired. To add, a particular piece which forms the piecewise curve can be easily modified without changing the shape of the whole piecewise Ball curve. Hence, one may form a fair curve [4,12] by inspecting the curvature plot. To conclude, this scheme yields a $G^2$ piecewise Ball curve which is local.

**Figure 1** illustrates an example of $G^2$ piecewise Ball curve rendered with the given interpolating points. The $G^2$ data used to construct the curve is as follows:

$$T_0 = (1,0),\ T_1 = (0,1),\ T_2 = (0,1) \text{ and } T_3 = (\frac{1}{\sqrt{2}}, \frac{1}{\sqrt{2}})$$

$$\kappa_0(0) = 3,\ \kappa_0(1) = \kappa_1(0) = 1,\ \kappa_1(1) = \kappa_2(0) = -1.5 \text{ and } \kappa_2(1) = -1$$

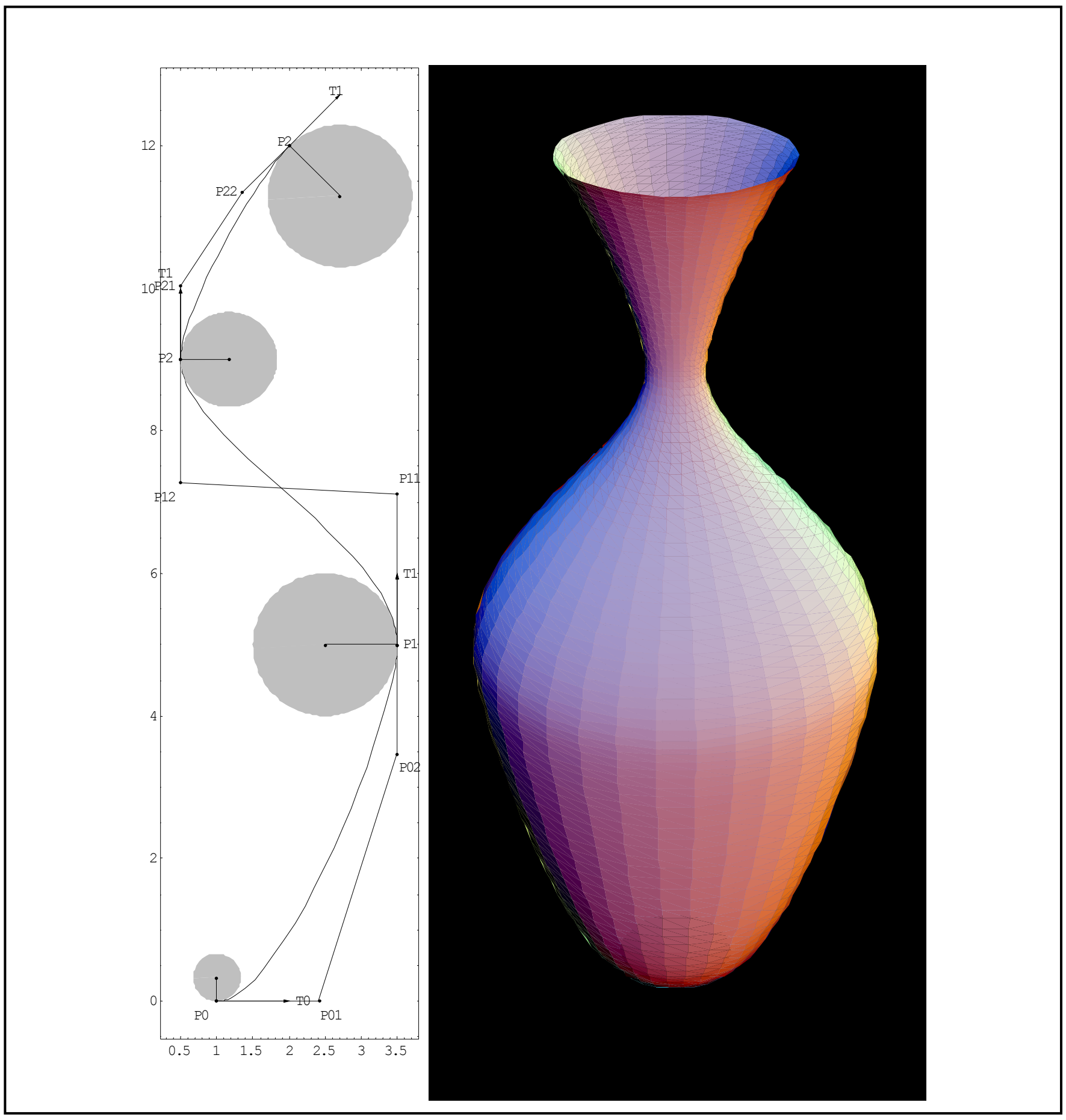

Figure 1: $G^2$ interactive algorithm yields a local piecewise Ball curve

## 5. Conclusion

In this paper, an interactive algorithm to generate $G^2$ piecewise Ball curve has been shown precisely. The strength of each scheme has been elaborated in the last section with regards to the profile of a vase. The proposed algorithm could be utilized by the designer to render a desired piecewise curve which has curvature continuity.

## References

[1]. A.A. Ball, CONSURF, Part 1: Introduction to conic title. Computer Aided Design, 6 (1974) 243-249.

[2]. A.A. Ball, CONSURF, Part 2: Description of the algorithms. Computer Aided Design, 7 (1975) 237-242.

[3]. A.A. Ball, CONSURF, Part 3: How the program is used. Computer Aided Design, 9 (1977) 9-12.

[4]. D.S. Meek, Coaxing a planar curve to comply, Journal of Computational and Applied Mathematics, 140 (2002) 599-618.

[5]. K. H Sollig and J. Koch, Geometric hermite interpolation with maximal order and smoothness, Comput. Aided Geom. Design, 13 (1996) 681–695.

[6]. T.N.T. Goodman and K. Unsworth, Shape preserving interpolation by curvature continuous parametric curves, Computer Aided Geometric Design, 5 (1988) 323-340.

[7]. R. Gobithasan, Designing geometrically continuous curves using Timmer parametric cubic, MSc. Thesis (2004) School of Mathematical Sciences, University Science Malaysia.

[8]. R. Gobithasan & M.A Jamaludin, Towards $G^2$ curve design using Timmer parametric cubic, Proceedings of International Conference of Computer Graphics, Imaging and Visualization, (CGIV04'), (2004) 109-114.

[9]. R. Gobithasan & M.A Jamaludin, Designing $G^2$ Timmer curves iteratively (In Malay), Proceedings of Seminar Masyarakat & Matematik, (SMM05'), (2004) 109-114.

[10]. H.B Said, Generalized Ball curve and its recursive algorithm, ACM. Trans. Graph., 8 (1989), 360-371.

[11]. T.N.T Goodman & H.B Said, Properties of generalized Ball curves and surfaces, Computer Aided Design, 23 (1991), 554-560.

[12]. G. Farin, Curvature and the fairness of curves and surfaces in IEEE, Computer Graphics & Applications. 9 (2) (1989b) 52-57.